\documentclass{pasj00}
\draft

\begin{document}
\SetRunningHead{Author(s) in page-head}{Running Head}
\Received{2007/01/09}
\Accepted{2007/02/09}

\title{CO Observations of a FeLoBAL Quasar with an H$\alpha$ Absorption
Line at $z=2.3$}

\author{
Kouji \textsc{Ohta}\altaffilmark{1},
Gaku \textsc{Kiuchi}\altaffilmark{1},
Kouichiro \textsc{Nakanishi}\altaffilmark{2},
Kentaro \textsc{Aoki}\altaffilmark{3},
Ikuru \textsc{Iwata}\altaffilmark{4}, \\
Masayuki \textsc{Akiyama}\altaffilmark{3}, 
Naoyuki \textsc{Tamura}\altaffilmark{3}, 
and
Masataka \textsc{Ando}\altaffilmark{1}}
\altaffiltext{1}{Department of Astronomy, Kyoto University, Kyoto 606-8502}
\email{ohta@kusastro.kyoto-u.ac.jp}
\altaffiltext{2}{Nobeyama Radio Observatory, Minamimaki Minamisaku, Nagano 384-1305}
\altaffiltext{3}{Subaru Telescope, National Astronomical Observatory of
Japan, 650 N. A'ohoku Place, Hilo, \\ HI 96720, USA}
\altaffiltext{4}{Okayama Astrophysical Observatory, National Astronomical Observatory, Okayama 719-0232}


%


\maketitle

\begin{abstract}
SDSS J083942.11+380526.3 is an Iron Low-ionization Broad Absorption Line
(FeLoBAL) quasar at $z = 2.3$, and Aoki et al. (2006) recently found 
the presence of an H$\alpha$ absorption line in the broad H$\alpha$ emission line.
Motivated by an idea that this quasar may be a huge molecular gas
reservoir in the early phase of quasar evolution, we made
CO($J=3-2$) observations of it using the Nobeyama Millimeter Array.
No significant CO emission was detected; although an emission-like feature
($2.5\sigma$) was seen close ($\sim 2''$) to the quasar, 
we regard it as a noise.
The obtained 3$\sigma$ upper limit on the CO luminosity is 
$L^{\prime}_{{\rm CO}(J=3-2)} = 4.5 \times 10^{10}$ K km s$^{-1}$ pc$^2$,
which corresponds to  $M({\rm H}_2) = 3.6 \times 10^{10} M_{\odot}$ if we
adopt the CO-to-H$_2$ conversion factor of 0.8 $M_{\odot}$ (K km
 s$^{-1}$ pc$^2)^{-1}$.
This upper limit is comparable to 
$L^{\prime}_{{\rm CO}(J=3-2)}$ (and thus the molecular gas mass) detected
in  quasars and BAL quasars at $z=1-3$, and 
no sign of the presence of the huge amount of molecular gas in this
FeLoBAL quasar was obtained. 
\end{abstract}

\section{Introduction}


FeLoBAL (iron LoBAL) quasars are a rare subclass of
broad absorption-line quasars (BAL quasars) recently
emerging thanks to the large sample of the Sloan Digital Sky
Survey (SDSS).
BAL quasars show broad (2,000 -- 20,000 km s$^{-1}$) blueshifted
(2,000 -- 60,000 km s$^{-1}$) absorption lines in the UV wavelength
region.
BAL quasars showing high-ionization absorption lines such as C IV, N V, Si IV,
and Ly$\alpha$ are called HiBAL quasars, and some BAL quasars also show 
low-ionization  absorption lines of Mg II, Al II, and Al III
in addition to the high-ionization absorption lines, 
which are called LoBAL quasars. 
About 10 -- 20 \% of optically selected quasars are BAL (HiBAL)
quasars and  $\sim 10$\% of them are LoBAL quasars (Weymann et al. 1991;
Hewett \& Foltz 2003; Reichard et al. 2003; Trump et al. 2006).
There is a subclass of LoBAL quasars which show numbers of 
absorption lines originated in Fe II and Fe III (Hazard et al. 1987; 
Cowie et al. 1994;  Becker et al. 1997, 2000; Reichard et al. 2003;
Hall et al. 2002).
They are called iron LoBAL (FeLoBAL) quasars and occupy about
15 \% of LoBAL quasars (Hall et al. 2002). 
 
Nature of BAL quasars is still not well understood.
There are two major views for BAL quasars: BAL quasars are non-BAL
quasars seen with a different viewing angle (e.g., Weymann et al. 1991;
Schmidt \& Hines 1999), and BAL quasars are young or recently 
refueled quasars (e.g., Boroson \& Meyers 1992; Becker et al. 2000).
In the latter view, the broad absorption lines appear when a nucleus is
blowing gas and dust out during a dust-enshroud quasar phase which
leads the BAL quasars to non-BAL quasars.
If this view is correct, BAL quasars (especially LoBAL quasars)
 are expected to also be ultra-luminous
infrared galaxies.
In fact, all four LoBAL and FeLoBAL quasars known to date at 
$z < 0.4$ 
have the infrared luminosities larger than $10^{12} L_{\odot}$
(Canalizo \& Stockton 2002).
Furthermore, Reichard et al. (2003) show that average UV spectrum
is getting flatter in the order of non-BAL, HiBAL, LoBAL, and FeLoBAL
quasars and the color excesses amount $E(B-V)\sim 0.023$ mag and $\sim
0.077$ mag for HiBAL and LoBAL quasars, 
respectively with Small Magellanic Cloud (SMC)-like reddening law.
The average UV spectrum of FeLoBAL quasars is much flatter than those
(Reichard et al. 2003)
and hence FeLoBAL quasars may be the most obscured ones among BAL quasars.
Thus a large amount of gas and dust may exist in LoBAL quasars,
especially in FeLoBAL quasars.

We found a FeLoBAL quasar at $z=2.3$ with an H$\alpha$ absorption line in 
the broad H$\alpha$ emission line.
The object, SDSS J083942.11+380526.3 (hereafter SDSS J0839+3805),
was originally found during our search for FeLoBAL quasars at $z=2.1 - 2.8$
by visual inspection of $\sim 4,800$ spectra  in the SDSS Data Release 3
(DR3; Abazajian et al. 2005).
The color excess of SDSS J0839+3805 is $E(B-V) \sim 0.15$ mag
with the SMC-like reddening law if the intrinsic spectrum is the 
same as a composite spectrum of non-BAL quasars, 
suggesting the presence of larger obscuration than typical LoBAL quasars.
Subsequent near-infrared spectroscopy with the Cooled Infrared 
Spectrograph and Camera
for OH Suppressor 
attached to the Subaru telescope
revealed the presence of H$\alpha$ absorption line in the broad
H$\alpha$ emission line (Aoki et al. 2006).
The absorption line is not originated in stellar continuum.
The absolute magnitude in the $i$-band is $-26.8$ mag (AB),
and this was the first discovery of such system in quasars;
only known such system was Seyfert 1 NGC 4151 at that time.
Very recently, two more FeLoBALs have been found to show  Balmer 
absorption lines (Hall 2007; Aoki et al. 2007).
The  width of the absorption line in SDSS J0839+3805 is $\sim 340$ km s$^{-1}$ 
with a blue shift of $\sim$ 520 km s$^{-1}$ relative to the H$\alpha$
emission line.
The equivalent width of the absorption line is estimated to be
$4.9-8.0$\AA\ depending on the employed intrinsic line profile of
the H$\alpha$ broad emission line.
The equivalent width  corresponds to a column density of neutral
hydrogen of $\sim 10^{18}$ cm$^{-2}$ if we assume a gas temperature
of $10^4$ K from the curve of growth (see Aoki et al. 2006 for details).
It suggests that SDSS J0839+3805 may be a large gas 
reservoir in the very young phase from a ultra-luminous infrared
galaxy to a quasar.
Motivated by these, we made CO observations of SDSS J0839+3805 aimed
at detecting the molecular gases.
In this paper, we adopt the flat  cosmology of
$H_0 = 70$ km s$^{-1}$ Mpc$^{-1}$, $\Omega_{\rm M}=0.3$, and
$\Omega_{\Lambda} =0.7$ .


\section{Observations}
CO ($J=3-2$) observations were made with the Nobeyama Millimeter Array 
(NMA) at the Nobeyama Radio Observatory
\footnote{The Nobeyama Radio Observatory is a branch of 
the National Astronomical Observatory of Japan, 
the National Institutes of Natural Sciences (NINS).}
from 2005 November 27 to December 1 with C-configuration  and 
from 2006 March 15 to March 18  and from 2006 April 3 to April 5
with D-configuration (more compact than C-configuration).
The expected observed frequency for the redshifted CO ($J=3-2$)
line is 104.221 GHz based on the redshift
determined with H$\alpha$ emission ($z=2.318\pm0.002$),
which would give the most reliable systemic redshift of the host galaxy.
The accuracy we achieved corresponds to a velocity uncertainty 
of only about 180 km s$^{-1}$, which is 
negligibly small for  the Ultra Wide Band Correlator 
(UWBC, Okumura et al. 2000) used  covering  1024 MHz 
($\sim 3000$ km s$^{-1}$).
In actual observations, however, we adopted the central frequency 
of 104.477 GHz to avoid the exact center of the spectrograph.
This displacement also does not cause a serious frequency coverage
problem.
A field of view was $\sim 65^{\prime\prime}$, and 
angular resolutions were 2$^{\prime\prime}$--3$^{\prime\prime}$ 
and 4$^{\prime\prime}$--6$^{\prime\prime}$
for C- and D-configuration, respectively.
A total integration time was 25 and 17 hours for the C- and the 
D-configuration, respectively.
The reference calibration was made with  radio source of B0821+394 for 
each observation. 
The flux calibration and band-pass calibration were made
with 3C273 ($22.54$ Jy at 105.7 GHz) 
for C-configuration in November 2005 and D-configuration 
in March 2006, and 3C84 ($6.43$ Jy at 110.2 GHz)
 for D-configuration in April 2006. 
The uncertainty in the flux scale calibrations was  estimated
to be $\sim10$\% for each observation.
The weather conditions were fine during the observing runs
with the C-configuration and the second run of the D-configuration.

The uv-data reduction was carried out by UVPROC-II package 
developed at the Nobeyama Radio Observatory (Tsutsumi et al. 1997).
After discarding bad visibilities and calibrating visibilities,
we produced channel maps and an integrated intensity map using 
NRAO AIPS package for each configuration.
The rms noises in the maps were 3.6 mJy beam$^{-1}$ and 5.0 mJy beam$^{-1}$
for C- and D-configuration, respectively with a velocity 
bin of 184 km s$^{-1}$.
Finally, we made the integrated intensity map using the combined data
taken with C- and D-configurations.
The synthesis beam size was $4^{\prime\prime}.0 \times
3^{\prime\prime}.0$
and the rms noise was 2.9 mJy beam$^{-1}$.

\section{Results and Discussion}
A contour map centered at 104.22 GHz integrated in the 184 km s$^{-1}$ velocity
bin  obtained with the C-configuration is shown in left panel of figure 1. 
A signal-like feature is seen 
at $\sim 2^{\prime\prime}$ southeast of the quasar position; 
the peak value in the map is 9 mJy beam$^{-1}$ with a significance of 
$\sim 2.5 \sigma$.
The width of the feature is about 300 km s$^{-1}$.
The feature was seen in each observing date with a worse significance level.
It should be noted that the position of the quasar is taken from SDSS catalog
and its accuracy is less than 0.$^{\prime\prime}$1
(Pier et al. 2003), and the accuracy of the astrometry in the NMA
observations is expected to be fairly less than $1^{\prime\prime}$.

However, in a contour map obtained with the D-configuration (the middle
panel of figure 1), the feature did not appear; if the feature is real
it corresponds to be 1.8$\sigma$ in the data obtained with 
D-configuration.
We made the map for each of the two observing runs with 
D-configuration, but we could not see the emission-like feature. 
Although  we examined the data taken with the C-configuration
and D-configuration carefully to see whether some spurious data exist or not, 
we  could not find such spurious data.

In the combined data (C- and D-configurations), 
the map integrated with the velocity bin of 184 km s$^{-1}$
centered at the 104.22 GHz is shown in the right panel of figure 1.
No emission feature with a significance larger
than 3$\sigma$ is present. 
The 2.5$\sigma$ feature is seen at about 2$^{\prime\prime}$ northeast of
the quasar position, but this position does not coincide with the
peak of the feature seen in the C-configuration. 
If the feature is real and we see the molecular gas in the
quasar, the displacement of $\sim2^{\prime\prime}$ of the feature 
might be  due to the low signal-to-noise ratio of the emission.
Alternatively, the displacement (a projected distance of
16 kpc at the redshift) might  be real.
A  double peaked distribution of molecular gas was not unusual
in high redshift quasars (Klamer et al. 2004);
for instance in the quasar BR 1202-0725 at $z=4.7$, no host galaxy has been
found for a  molecular gas component separated by 4$^{\prime\prime}$
(projected distance of 26 kpc at the redshift)
from the quasar position (Carilli et al. 2002).
Thus there might be a case that the molecular gas in the quasar
itself is not rich, while it is gas rich in a putative close companion.
In any cases, however, the feature only shows the $2.5 \sigma$ significance
at most.
The feature is not seen in the D-configuration data and 
the location seen in the combined data is different from that 
seen in the C-configuration data.
Therefore we regard that this feature is not real and derive 
the upper limit on the molecular gas mass below.

From the rms noise, we set the upper limit on the CO luminosity 
as well as the molecular gas mass as follows,
$$ L_{{\rm CO} (J=3-2)}^{\prime} = 
{c^2 \over 2k} d_L^2 3 \sigma \Delta v \nu_{\rm rest}^{-2} (1+z)^{-1} \\ 
= 4.5 \times 10^{10}\ {\rm K\ km\ s}^{-1}\ {\rm pc}^2,$$
where, $c$ is the light speed, $k$ is the Boltzmann constant, $d_L$
is the luminosity distance, $\sigma$ is the rms noise (2.9 mJy
beam$^{-1}$) of the combined data at the velocity width of
$\Delta v$ (184 km s$^{-1}$), 
$\nu_{\rm rest}$ is the rest frequency of the target line.
In order to estimate the molecular gas mass, the conversion factor from
$L_{\rm CO}^{\prime}$ to $M({\rm H}_2)$, $\alpha$, is necessary.
Although the value of $\alpha$ in this system is unknown, we adopt
$\alpha =0.8\ M_{\odot}$ (K km s$^{-1}$ pc$^2$)$^{-1}$ to compare 
recent results of the CO observations of high redshift quasars
(Solomon \& Vanden Bout 2005).
Here we also assume that $L^{\prime}_{{\rm CO}(J=3-2)}$ is equal to
 $L^{\prime}_{{\rm CO}(J=1-0)}$ .
The resulting upper limit on the molecular gas mass is 
$3.6 \times 10^{10} M_{\odot}$.

\begin{figure}  \begin{center}
    \FigureFile(170mm,80mm){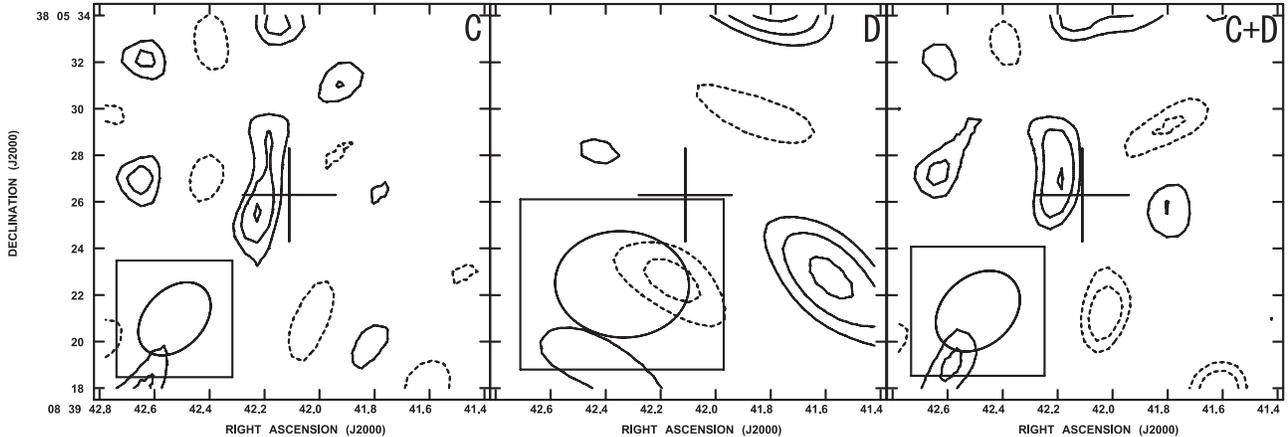}
  \end{center}
  \caption{Integrated CO($J=3-2$) map toward SDSS J0839+3805 
at observed frequency of 104.22 GHz with a 184 km s$^{-1}$ velocity width.
Contour levels are $-2$, $-1.5$, 1.5, 2, 2.5 $\sigma$ with 
 $1 \sigma = 3.6$ mJy beam$^{-1}$ in the C-configuration (left panel), 
 $1 \sigma = 5.0$ mJy beam$^{-1}$ in the D-configuration(middle panel), 
 and $1 \sigma = 2.9$ mJy beam$^{-1}$ in the C$+$D(right panel). Cross at the
 center shows the position of the FeLoBAL quasar. Beamsize is shown in
 the bottom-left corner of  each panel.}\label{fig:sample}
\end{figure}

The upper limit on the molecular gas mass of $3.6 \times 10^{10} M_{\odot}$ 
(or $L^{\prime}_{{\rm CO}(J=3-2)}$ of $4.5 \times 10^{10}$ K km s$^{-1}$
pc$^{2}$) should be confronted with the molecular gas masses 
detected in quasars at $z=1-3$;
they are mostly derived from $L^{\prime}_{{\rm CO} (J=3-2)}$ and
 range from $0.4 \times 10^{10} M_{\odot}$ to 
$7.4 \mu^{-1} \times 10^{10} M_{\odot}$ 
(Solomon and Vanden Bout 2005, with the same
cosmology and the conversion factor), where $\mu$ refers to the 
magnification factor of the gravitational lens.
The largest value comes from a very red quasar presumably obscured by
dust (MG0414+0534).
Among them, Cloverleaf (H1413+117) and VCV J1409+5628 are BAL quasars, and their molecular
gas masses are $3.2 \times 10^{10} M_{\odot}$ (delensed with $\mu=11$) 
and $6.3 \mu^{-1} \times 10^{10} M_{\odot}$, respectively.
Therefore  the FeLoBAL of SDSS J0839+3805 is not very molecular gas rich
as compared with quasars and BAL quasars at the similar redshifts.
This is true even if the emission-like feature mentioned above is
real.


In conclusion, our target of FeLoBAL SDSS J0839+3805 with the Balmer
absorption line is turned out not to be  very molecular gas rich among 
quasars at the similar redshift, and it is the case even 
if the emission-like feature is real.
However  this is the first trial observation of a FeLoBAL quasar 
in our knowledge, further observations in submillimeter 
(Lewis et al. 2003; Willott et al. 2003; Priddey et al. 2006) and 
CO line  are necessary to examine the amount of molecular gases
and its relation to the line-of-sight absorption
 in FeLoBAL quasars as well as BAL quasars.

\vspace{1.0cm}
We thank the staff members of the Nobeyama Radio Observatory
for the help of the observations.
KO acknowledges the support by Grant-in-Aid for Scientific Research
from Japan Society for Promotion of Science (17540216).
















\end{document}